\documentclass[12pt]{revtex4}
\usepackage{pslatex}
\usepackage[T1]{fontenc}
\usepackage{epsf}

\begin{document}
\title{Energy difference between the lowest doublet and quartet
states of the boron atom}
\author{Krzysztof Strasburger}
\affiliation{Department of Physical and Quantum Chemistry \\
Faculty of Chemistry \\
Wroc{\l}aw University of Science and Technology \\
Wybrze\.{z}e Wyspia\'{n}skiego 27, 50-370 Wroc{\l}aw, Poland}
\date{\today}

\begin{abstract}
The energies of the lowest $^2P_u$, $^4P_g$ and $2D_g$ states of the boron atom
are calculated with $\mu$hartree accuracy, in the basis of symmetrized,
explicitly correlated Gaussian lobe functions. Finite nuclear mass and scalar
relativistic corrections are taken into account.
This study contributes to the problem of the energy differences
between doublet and quartet states of boron, which have not been measured
to date. It is found that the $^2P_u\rightarrow^4P_g$ excitation energy,
recommended in the Atomic Spectra Database, appears
underestimated by more than 300~cm$^{-1}$.
\end{abstract}
\maketitle

\section{Introduction}

Highly accurate calculations, carried out within the well-grounded theory
of quantum mechanics, are currently possible for few-electron atoms and
molecules. The results are usually compared with spectroscopic data.
This collation verifies the theory
and computational methods, but may also stimulate improvements of the
experiment. History of the studies on the rovibrational spectrum of the
hydrogen molecule is a good example of such positive feedback \cite{H2-1,H2-2}.
The calculations may also provide reliable results where experimental data
are missing. For the boron atom, intersystem radiative transitions
were not observed, therefore the energy differences between
the spin doublet and quartet states, listed in the Atomic Spectra Database
(ASD) \cite{ASD}
are based on numerical extrapolation of the transition energies
known for heavier, isoelectronic ions \cite{Kramida}. 

According to this extrapolation, the lowest $^4P_g$ term has the energy higher
by 28644.3 cm$^{-1}$, than the ground state term ($^2P_u$).
The $J$ quantum number is omitted, because the fine structure is not considered
in the present work. The energy of a non-splitted term is not observable,
and is computed from experimental data, as
weighted average over associated,
$J$-dependent term energies. Calculations of this energy difference were also
carried out in the past, but the results
do not agree with that ``experimental'' value. The short review is
limited to most recent articles, because the results of earlier calculations
\cite{Jonsson,Galvez} were simply
too innaccurate for a comparison with spectroscopic data.
Froese Fischer and coworkers \cite{CFF}
used the multiconfiguration Hartree-Fock (MCHF) method, with finite nuclear
mass and scalar relativistic corrections taken into account, and obtained
the excitation energy amounting to 28959(5)~cm$^{-1}$.
It is to be noted that their computational method was
validated for the carbon cation, with theoretical result different from
experimental one by only 7~cm$^{-1}$.
Chen \cite{Chen} predicted 28719.46~cm$^{-1}$,
using Configuration Interaction wave function, and also including
relativistic and finite nuclear mass corrections. Nakatsuji and coworkers
\cite{FC-Nakatsuji} employed the free-complement chemical-formula-theory
(FC-CFT) method. The value of
28826~cm$^{-1}$ is obtained, with their nonrelativistic, fixed-nucleus
energies, and assuming that the respective corrections would
contribute c.a. 50~cm$^{-1}$, similarly as in the calculations by Chen
and Froese Fischer. The largest discrepancy between theoretical and
experimental excitation energy exceeds 300~cm$^{-1}$. Computational results
are however rather scattered
and a decisive calculation requires a wave function that provides sufficiently
accurate absolute electronic energies. Apart of the $^2P_u$ and $^4P_g$
states, the lowest $^2D_g$ state is also the subject of the present study,
because the transition energies to the latter, from the ground state, are
known and may serve for estimation of uncertainty of final results.
Experiment-based energy difference between $^2P_u$ and $^2D_g$ terms amounts to
47846.74~cm$^{-1}$ \cite{ASD}.

In theoretical studies of the boron atom, not necessarily aimed at the
$^2P_u\rightarrow^4P_g$ excitation, most efforts to date
were devoted to the ground state \cite{Sasaki,Feller,Meyer}.
Preliminary Hylleraas-CI calculations
were reported by Ruiz \cite{B-Ruiz}. Highly accurate, nonrelativistic
energies were obtained with the explicitly correlated
r$_{12}$-MR-CI method \cite{Gdanitz}, and in the diffusion Monte Carlo (DMC)
simulations \cite{atoms-Seth}. There is a masterpiece of CI
calculations, by Almora-Diaz and Bunge \cite{B-Almora-Diaz}, with the orbital
basis containing functions corresponding to the $l$ quantum number reaching 20
($z$-type orbitals), yielding the
energy only 31 $\mu$hartree above the variational limit. Well-hit
extrapolation to complete basis set missed this limit by 6$\mu$hartree. 
Best results were obtained with explicitly correlated Gaussian functions (ECG)
\cite{B-Bubin,B-Puchalski}. The estimated error of nonrelativistic energy
of this state was smaller than 1 $\mu$hartree. Similar accuracy was
achieved for the $^2S_g$ states, and the transition energies between
the ground state and $S$-symmetry states were reproduced within a fraction
of cm$^{-1}$, with finite nuclear mass, relativistic (including fine and
hyperfine structure for the ground state term) and leading radiative
corrections taken into account.

The wavefunctions and energies of comparable accuracy are missing for the
$^4P_g$ and $^2D_g$ states, and the results are scarce in the literature
\cite{Jonsson,Galvez,FC-Nakatsuji,CFF}. 
The present paper is aimed at filling in this hole, and contributing to
final resolution of the discrepancies concerning the energy differences
between the spin doublet and quartet terms of the boron atom.

Nonrelativistic wavefunctions, expressed as linear combinations of
symmetry-adapted, explicitly correlated Gaussian functions, and variational
energies with scalar relativistic corrections are obtained for the lowest
$^2P_u$, $^4P_g$, and $^2D_g$ states.
Atomic units are used unless stated otherwise. Conversion factor to the
energy unit used commonly in spectroscopy amounts to
1~hartree=219474.63~cm$^{-1}$

\section{Method}

The stationary Schr\"{o}dinger equation for $n$-electron atom is solved with
the nonrelativistic Hamiltonian
\begin{equation}
\hat{H}=-\frac{\nabla^2_{nuc}}{2m_{nuc}}+\sum_{i=1}^{n}\left(-\frac{\nabla^2_i}{2}
-\frac{Z}{r_i}\right)+\sum_{i>j=1}^{n}\frac{1}{r_{ij}}
\label{nrham}
\end{equation}
where $i$ and $j$ count the electrons. 
Details of the method  have been introduced
in earlier papers devoted to the lithium and carbon atoms
\cite{Li-KS,C-KS}, and various states
of many-electron harmonium \cite{harm3-JC,harm4-JC,harm56-JC,harm6-KS}.
The wavefunction
\begin{equation}
\Psi({\bf r}_1,s_1,\ldots,{\bf r}_n,s_n)=\sum_{I=1}^K C_I \hat{A}\Theta_I(s_1,\ldots,s_n)\hat{P}\chi_I({\bf r}_1,\ldots,{\bf r}_n)
\label{wavefn}
\end{equation}
is expressed as linear combination
of explicitly correlated Gaussian primitives (lobes)
\begin{equation}
\chi_I({\bf r}_1,\ldots,{\bf r}_n)=\exp{\left[-\sum_{i=1}^na_{I,i}({\bf r}_i-{\bf R}_{I,i})^2-\sum_{i>j=1}^nb_{I,ij}r^2_{ij}\right]},
\label{primit}
\end{equation}
symmetrized by the spatial symmetry projector $\hat{P}$, proper for chosen
one-dimensional, irreducible representation of selected finite point group.
This wavefunction is not an eigenfunction of the square of angular momentum
operator ($\hat{L^2}$), for non-zero ${\bf R}_{I,i}$ vectors. The deviation
from exact $L(L+1)$ eigenvalue is effectively diminished by the procedure of
variational energy minimization, in which the parameters (linear $C_I$
and nonlinear $a_{I,i}$, $b_{I,ij}$, and ${\bf R}_{I,i}$)
are established. Action of $\hat{P}$ upon $\chi_I$ annihilates from the
wavefunction, a finite subset of unwanted components, whose symmetry properties
are specific to some other representations of the $K_h$ point group,
and ensures convergence towards desired state.
$\Theta_I(s_1,\ldots,s_n)$ is the spin function, common for all basis functions
for given state, which is sufficient, because the spatial functions are
nonorthogonal. Namely,
\begin{equation}
\Theta_I(s_1,\ldots,s_5)=[\alpha(1)\beta(2)-\beta(1)\alpha(2)][\alpha(3)\beta(4)-\beta(3)\alpha(4)]\alpha(5)
\end{equation}
is used for both doublets, and
\begin{equation}
\Theta_I(s_1,\ldots,s_5)=[\alpha(1)\beta(2)-\beta(1)\alpha(2)]\alpha(3)\alpha(4)\alpha(5)
\end{equation}
for the quartet. $\hat{A}$ is the antisymmetrizer, which ensures proper
permutational symmetry of the wavefunction.

The relativistic energy of a resting system may be written as the power
series of the fine structure constant
$\alpha=\frac{1}{4\pi\epsilon_0}\frac{e^2}{\hbar c}$. Omitting the rest mass
contribution, 
\begin{equation}
E_{rel}=E_{nr}+E^{(2)}+E^{(3)}+\cdots
\end{equation}
where $E_{nr}$ is the nonrelativistic energy, $E^{(2)}$ contains the
Breit-Pauli relativistic corrections and higher order terms are known
as the radiative (QED) corrections. All these corrections may be calculated
in perturbative manner, as expectation values of
respective operators, with known nonrelativistic wavefunction.
The Breit-Pauli Hamiltonian may be split to the relativistic shift
$\hat{H}_{RS}$ operator, with expectation value $E_{RS}$,
and the fine and hyperfine structure operators,
which contain spin-orbit and spin-spin coupling terms.
Only the former is considered in this work. It is convenient to write
it down as the sum of following terms:
\begin{equation}
\hat{H}_{RS}=\hat{H}_1+\hat{H}_{1n}+\hat{H}_2+\hat{H}_3+\hat{H}_4+\hat{H}_{4n}.
\label{RS}
\end{equation}
These operators describe respectively the electronic mass-velocity correction
\begin{equation}
\hat{H}_{1}=-\frac{1}{8c^2}\sum_{i=1}^n\nabla^4_i,
\end{equation}
the electron-nucleus Darwin term
\begin{equation}
\hat{H}_2=\frac{Z\pi}{2c^2}\sum_{i=1}^n\delta({\bf r}_i),
\end{equation}
the sum of the electron-electron Darwin term and spin-spin Fermi
contact interaction (both have the same mathematical form, after integration
over spin variables \cite{Davidson-CH2})
\begin{equation}
\hat{H}_3=\frac{\pi}{c^2}\sum_{i>j=1}^n\delta({\bf r}_{ij}),
\end{equation}
and the electron orbit-orbit term
\begin{equation}
\hat{H}_4=\frac{1}{2c^2}\sum_{i>j=1}^n\left(\frac{\nabla_i\cdot\nabla_j}{r_{ij}}+
\frac{{\bf r}_{ij}\cdot[({\bf r}_{ij}\cdot\nabla_i)\nabla_j]}{r_{ij}^3}\right),
\end{equation}
which describes the interaction of magnetic dipoles arising from orbital motion
of the electrons. There are two terms in equation \ref{RS}, that have non-zero
value only for finite nuclear mass, namely the nuclear
mass-velocity correction
\begin{equation}
\hat{H}_{1n}=-\frac{1}{8m_{nuc}^3c^2}\nabla^4_{nuc},
\end{equation}
and the nucleus-electron contribution to
orbit-orbit magnetic interaction energy
\begin{equation}
\hat{H}_{4n}=-\frac{Z}{2m_{nuc}c^2}\sum_{i=1}^n\left(\frac{\nabla_i\cdot\nabla_{nuc}}{r_{ij}}+
\frac{{\bf r}_{i}\cdot[({\bf r}_{i}\cdot\nabla_i)\nabla_{nuc}]}{r_{i}^3}\right).
\end{equation}

Distinction of the cases of fixed and non-fixed nucleus requires only the
modification of the nuclear mass in all Hamiltonians, from infinity
to the one proper for given isotope of boron. The wavefunction given by Eqs.
\ref{wavefn} and \ref{primit} is expressed
in relative coordinates --- ${\bf r}_i$ denotes the position of i$^{th}$
electron relatively to the nucleus. 
Therefore explicit transformation of the operators, both nonrelativistic and
relativistic, from laboratory to center-of-mass coordinate frame, is not
necessary. Only relative coordinates appear in these operators explicitly.
Each differrentiation over a coordinate in Cartesian laboratory frame, may be
written as properly weighted sum of differentiations over respective
relative and center-of-mass coordinates. Differentiation of a function,
which is dependent on relative coordinates only, over a center-of-mass coordinate,
gives zero, so the final result is the same with non-transformed operators
as with explicit elimination of the center of mass motion \cite{BHA-Kutz}.

\section{Numerical results}

In the first step, nonrelativistic wavefunctions are
constructed.
The ground state wavefunction of the boron atom has $P_u$ symmetry. Assuming
the magnetic quantum number equal to 0, this symmetry is
effectively represented by the $A_u$ representation of the $C_i$ point group,
with the projector
\begin{equation}
\hat{P}=\hat{E}-\hat{i}
\end{equation}
and all ${\bf R}_{I,i}$ vectors placed at the $z$-axis of the coordinate
frame. The $C_{4v}$ point group is employed for both excited
states, with ${\bf R}_{I,i}$ vectors confined to the $xy$ plane.
The projector proper for the $A_2$ representation,
\begin{equation}
\hat{P}=\hat{E}+\hat{C}_4^1+\hat{C}_2+\hat{C}_4^3-\hat{\sigma}_{v1}-\hat{\sigma}_{v2}-\hat{\sigma}_{d1}-\hat{\sigma}_{d2}
\end{equation}
produces effectively the $P_g$ symmetry of the quartet state, and the $B_1$
representation, with
\begin{equation}
\hat{P}=\hat{E}-\hat{C}_4^1+\hat{C}_2-\hat{C}_4^3+\hat{\sigma}_{v1}+\hat{\sigma}_{v2}-\hat{\sigma}_{d1}-\hat{\sigma}_{d2}
\end{equation}
is adequate for the $D_g$ state, producing the wavefunction converging to
the normalized sum of eigenfunctions of $\hat{L}_z$, pertaining to $m_L=2$
and $m_L=-2$.

\begin{table}
\caption{Nonrelativistic energies, deviations of $<L^2>$ from $L(L-1)$, and
extrapolated energies, for fixed nucleus. For extrapolated ($E_{extr}$) results,
standard deviations of the least significant digits are given in parentheses
\label{tab-fixed-nonrel}}
\begin{tabular}{|r|c|c||r|c|c|} \hline
 K  & $E_{nr}$ & $\langle L^2\rangle-L(L+1)$ &  K  & $E_{nr}$ & $\langle L^2\rangle-L(L+1)$  \\ \hline
\multicolumn{6}{|l|}{$^2P_u$ (L=1)} \\ \hline
277   & $-24.653001970$ & $7.81\cdot 10^{-6}$ & 2745  & $-24.653862346$ & $1.73\cdot 10^{-7}$ \\
406   & $-24.653462344$ & $5.76\cdot 10^{-6}$ & 4022  & $-24.653865404$ & $8.82\cdot 10^{-8}$ \\
595   & $-24.653681184$ & $3.48\cdot 10^{-6}$ & 5679  & $-24.653867017$ & $4.97\cdot 10^{-8}$ \\
872   & $-24.653785377$ & $1.83\cdot 10^{-6}$ & 7456  & $-24.653867660$ & $3.54\cdot 10^{-8}$ \\
1278  & $-24.653833991$ & $7.33\cdot 10^{-7}$ & 10304 & $-24.653868064$ & $2.12\cdot 10^{-8}$ \\
1873  & $-24.653854171$ & $3.30\cdot 10^{-7}$ & $E_{extr}$ & $-24.65386890(14)$ & 0 \\ \hline
\multicolumn{6}{|l|}{$^4P_g$ (L=1)} \\ \hline
277  & $-24.521826756$ & $1.10\cdot 10^{-5}$ & 1873 & $-24.522039020$ & $2.26\cdot 10^{-7}$ \\
406  & $-24.521944458$ & $7.17\cdot 10^{-6}$ & 2733 & $-24.522040459$ & $1.11\cdot 10^{-7}$ \\
595  & $-24.521999781$ & $3.07\cdot 10^{-6}$ & 3580 & $-24.522041147$ & $5.47\cdot 10^{-8}$ \\
872  & $-24.522023448$ & $1.47\cdot 10^{-6}$ & 4672 & $-24.522041430$ & $3.49\cdot 10^{-8}$ \\
1278 & $-24.522035395$ & $4.74\cdot 10^{-7}$ & $E_{extr}$ & $-24.52204180(5)$ & 0 \\ \hline
\multicolumn{6}{|l|}{$^2D_g$ (L=2)} \\ \hline
277  & $-24.434439490$ & $1.86\cdot 10^{-4}$ & 2745 & $-24.435961389$ & $5.34\cdot 10^{-6}$ \\
406  & $-24.435110865$ & $1.43\cdot 10^{-4}$ & 4023 & $-24.435972976$ & $2.69\cdot 10^{-6}$ \\
595  & $-24.435568403$ & $7.98\cdot 10^{-5}$ & 5858 & $-24.435978480$ & $1.30\cdot 10^{-6}$ \\
872  & $-24.435789658$ & $4.04\cdot 10^{-5}$ & 8231 & $-24.435981009$ & $5.35\cdot 10^{-7}$ \\
1278 & $-24.435896508$ & $1.90\cdot 10^{-5}$ & & & \\
1873 & $-24.435941219$ & $1.01\cdot 10^{-5}$ & $E_{extr}$ & $-24.43598347(63)$ & 0 \\ \hline
\end{tabular}
\end{table}

\begin{figure}
\caption{Energy extrapolation using deviation of $\langle L^2\rangle$ from
L(L+1), for the $^2P_u$ state \label{fig-extrap-2Pu}}
\epsfxsize=12cm \epsfbox{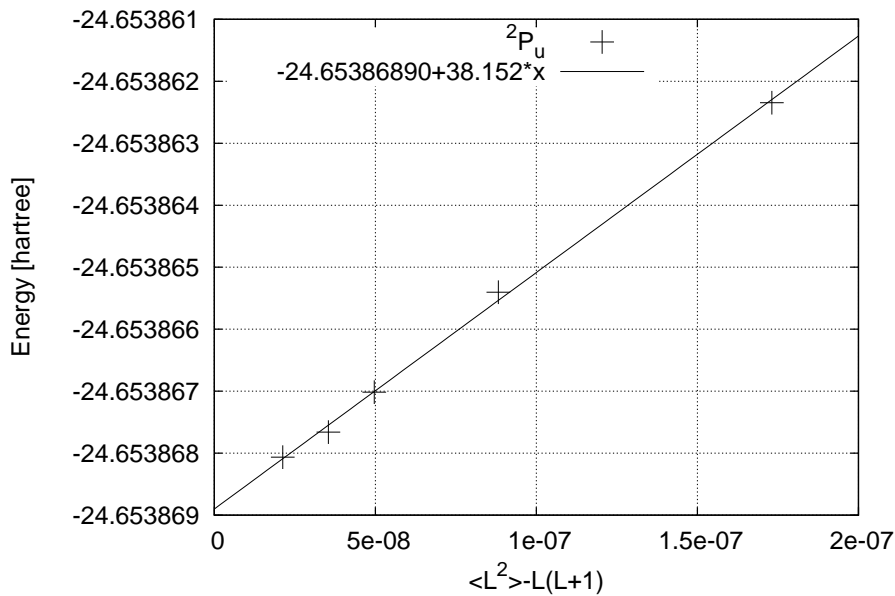}
\end{figure}

\begin{figure}
\caption{Energy extrapolation using deviation of $\langle L^2\rangle$ from
L(L+1), for the $^4P_g$ state \label{fig-extrap-4Pg}}
\epsfxsize=12cm \epsfbox{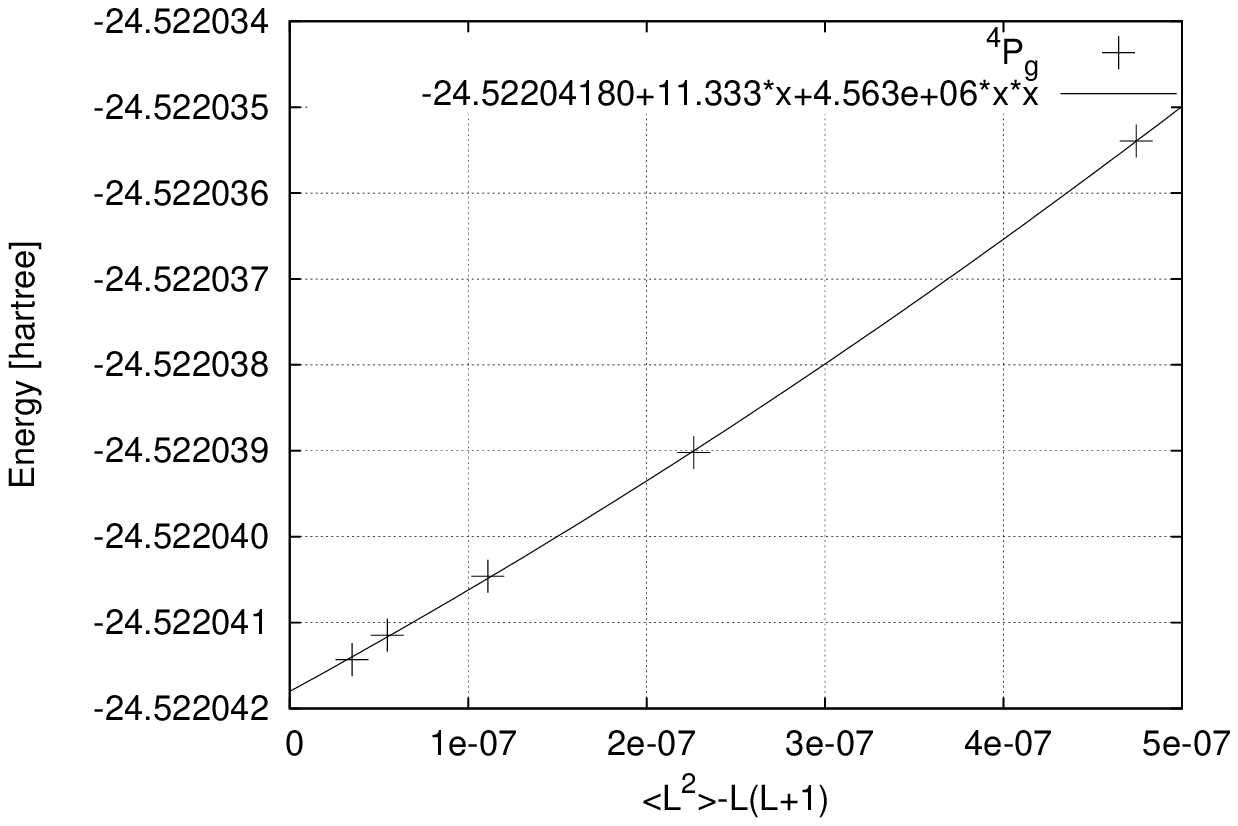}
\end{figure}

\begin{figure}
\caption{Energy extrapolation using deviation of $\langle L^2\rangle$ from
L(L+1), for the $^2D_g$ state \label{fig-extrap-2Dg}}
\epsfxsize=12cm \epsfbox{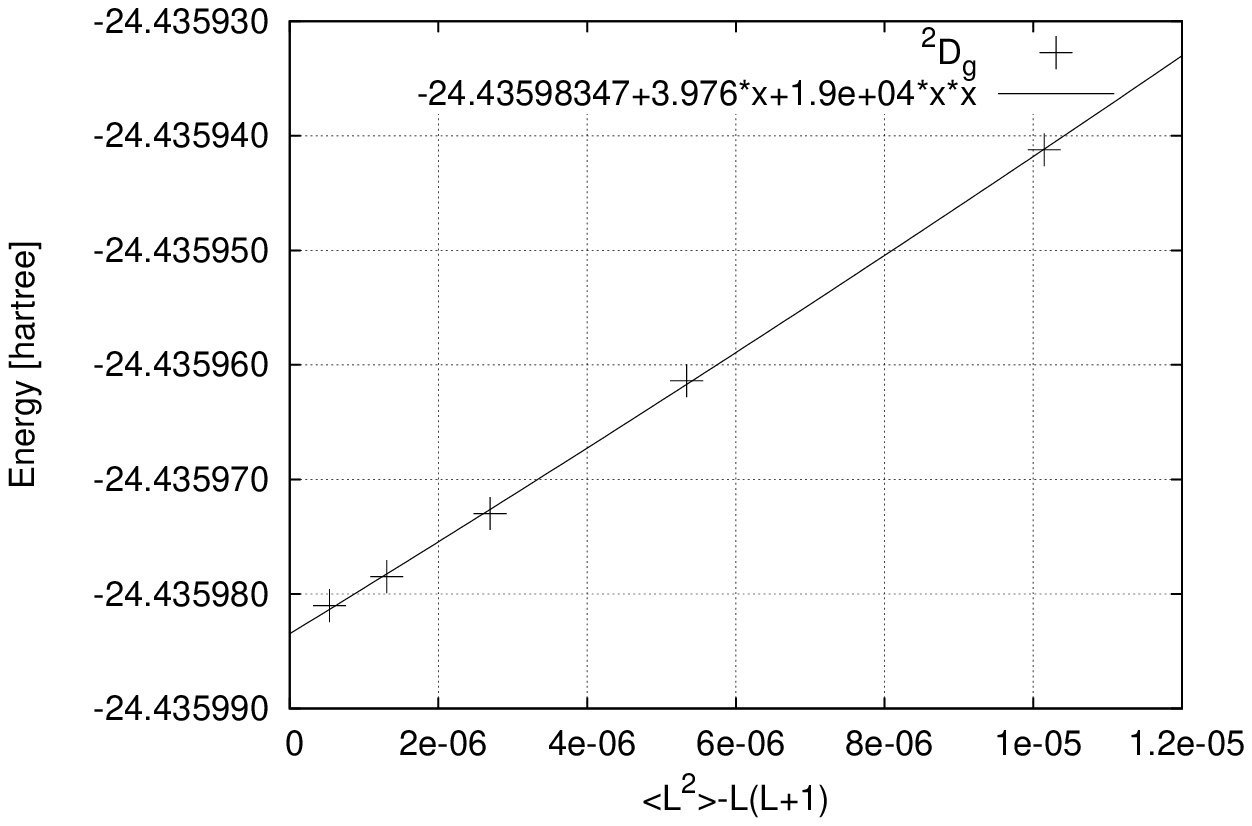}
\end{figure}

The accuracy of
nonrelativistic energies is assessed, exploiting the convergence of
$\langle\hat{L}^2\rangle$, whose known exact limits amount to $L(L+1)$.
Basis sets were extended stepwise, beginning with 1, 2 and 3 ECGs and
then appending functions optimized two steps
back in the process, to the current set. Optimization of all
variational parameters of the new basis followed, aimed at energy minimization.
Successive basis sizes formed thus initially the Narayana's cows sequence
\cite{integers}. For large bases, functions appeared that contributed too
little to the energy, and these functions were removed from the set.
The threshold value was set to 1, $0.5$ or $0.2$ nanohartree, dependent
on the estimated distance to the variational limit. The values of nonrelativistic
energies and $\langle L^2\rangle$, calculated for infinite-mass nucleus,
with $K$ basis functions, are collected in table \ref{tab-fixed-nonrel}.
It is noticed that the energy depends smoothly on the error of the square of
angular momentum, $\langle L^2\rangle-L(L+1)$ --- similarly as for the
carbon atom \cite{C-KS}. This observation, which
has no theoretical background and may be related to the method of
construction of consecutive basis sets, gives rise to an assumption that
the rotational energy error becomes nearly constant fraction of the total
energy error. Either linear (for the ground state, Fig. \ref{fig-extrap-2Pu})
or quadratic (for both excited states, Figs. \ref{fig-extrap-4Pg} and
\ref{fig-extrap-2Dg}) functions are fitted to five best points, giving
estimations of complete basis set limits of the electronic energies.
Variational energies look converged to a fraction of
$\mu$hartree for $^2P_u$ and $^4P_g$ states, while the accuracy for the
$^2D_g$ state is a little worse, with the distance to the estimated limit still
amounting to c.a. $2.5 \mu$hartree. The wavefunction of this state has
apparently more complicated character, but calculation with a significantly
larger basis set was not feasible.

\begin{table}
\caption{Comparison of nonrelativistic energies with published results
\label{complit}}
\begin{tabular}{|l|l|l|l|} \hline
method & $^2P_u$ & $^4P_g$ & $^2D_g$ \\ \hline
MCHF ($l_{max}=7$) \cite{Jonsson} & $-24.651009$ & & $-24.431353$ \\
VMC \cite{Galvez}  & $-24.64502(6)$ & $-24.51581(6)$ & $-24.42486(5)$ \\
CI ($l_{max}=6$, selected) \cite{Chen} & $-24.652032$ & $-24.521401$ & $-24.433575$ \\
MCHF ($l_{max}=5$) \cite{CFF} & $-24.653523595$ & $-24.521822334$ & \\
FC-CFT \cite{FC-Nakatsuji} & $-24.653734(103)$  & $-24.522622(50)$ & \\
r$_{12}$-MR-CI \cite{Gdanitz} & $-24.653787$ & & \\
DMC \cite{atoms-Seth} & $-24.65379(3)$ & & \\
CI ($l_{max}=20$) \cite{B-Almora-Diaz} & $-24.65383733$  & & \\
CI, extrapolated \cite{B-Almora-Diaz} & $-24.653862(2)$ & & \\
ECG, K=5100 \cite{B-Bubin}  & $-24.65386608$ & & \\
ECG, K=8192 \cite{B-Puchalski} & $-24.653867537$ & & \\
ECG, extrapolated \cite{B-Puchalski} & $-24.65386805(45)$ & & \\
ECG lobes (present work) & $-24.653868064$  & $-24.522041430$ & $-24.435981009$ \\
$E_{extr}$ (present work) & $-24.65386890(14)$ & $-24.52204180(5)$ & $-24.43598347(63)$ \\ \hline
\end{tabular}
\end{table}

Comparison with literature data, in table \ref{complit}, reveals that
the variational
energy of the ground state, obtained in the present work with 7456 basis
functions, is lower than the best previous result \cite{B-Puchalski} by
$0.5$~$\mu$hartree, and with 10304 basis functions surpasses also
the old estimate of the complete basis set limit.
The $\langle L^2\rangle$-based extrapolation lowers this limit by $0.85
\mu$hartree. There
are no published energies of comparable accuracies, for both excited
states. The calculation by Nakatsuji \cite{FC-Nakatsuji} yielded the energy
of the ground state, higher by 0.135 mhartree than the present result. On
the contrary, the energy of the $^4P_g$ state was too low, overstepping
the variational limit by 0.58 mhartree.
The MCHF energies by Froese Fischer \cite{CFF} look more balanced, being
higher by $0.345$ ($^2P_u$ state) and $0.219$ ($^4P_g$ state) mhartree.
Most accurate nonrelativistic energy of the $^2D_g$ state, published to date
\cite{Chen}, is by more than 2 mhartree higher than the present one.

Concerning the components of relativistic corrections
(table \ref{tab-fixed-rs}), the convergence of the mass-velocity
and electron-nucleus Darwin terms is still unsatisfactory for all states,
with differences of few
$\mu$hartree, between two most accurate wavefunctions. This inaccuracy
is due to $\nabla^4$ and $\delta({\bf r})$ operators, whose expectation values
converge very slowly in the basis of Gaussian functions, which do not represent
properly the wavefunctions at coalescence points (cusps). Fortunately, the
errors of $\langle\hat{H}_1\rangle$ and $\langle\hat{H}_2\rangle$ have
opposite signs and cancel to a significant extent. The number of stable
significant digits of $\langle\hat{H}_3\rangle$ is even smaller than that of
$\langle\hat{H}_2\rangle$, but the absolute value is smaller by two orders
of magnitude. On the other hand, the orbit-orbit
magnetic interaction energies look accurate
within one nanohartree. Total relativistic corrections (last column of table 
\ref{tab-fixed-rs}), calculated with two largest basis sets, differ by less
than $0.1$ $\mu$hartree for all states, although there is
no way to extrapolate these results and estimate the error margin more
rigorously. For the ground state, the results by Puchalski \cite{B-Puchalski}
are available, obtained with the method that involves regularization
of the $\nabla^4$ and $\delta({\bf r})$ operators, which leads to
much better convergence, and yields the scalar relativistic correction
amounting to $-7.515977$ mhartree. This means that the error of best present
calculation amounts to $0.141$ $\mu$hartree.

\begin{table}
\caption{Scalar relativistic corrections (in mhartree),
for fixed nucleus \label{tab-fixed-rs}}
\begin{tabular}{|r|c|c|c|c|c|} \hline
 K  & $\langle\hat{H}_1\rangle$ & $\langle\hat{H}_2\rangle$ & $\langle\hat{H}_3\rangle$ &
$\langle\hat{H}_4\rangle$ & $E_{RS}$ \\ \hline
\multicolumn{6}{|l|}{$^2P_u$} \\ \hline
277   & $-36.599999$ & $29.743913$ & $-0.601678$ & $-0.057897$ & $-7.515662$ \\
406   & $-36.728411$ & $29.867256$ & $-0.597815$ & $-0.057872$ & $-7.516842$ \\
595   & $-36.806576$ & $29.938863$ & $-0.595500$ & $-0.057843$ & $-7.521055$ \\
872   & $-36.834510$ & $29.969803$ & $-0.594004$ & $-0.057833$ & $-7.516544$ \\
1278  & $-36.863671$ & $29.998464$ & $-0.593480$ & $-0.057827$ & $-7.516515$ \\
1873  & $-36.873790$ & $30.008434$ & $-0.592950$ & $-0.057823$ & $-7.516129$ \\
2745  & $-36.893141$ & $30.027586$ & $-0.592638$ & $-0.057822$ & $-7.516015$ \\
4022  & $-36.897900$ & $30.032285$ & $-0.592543$ & $-0.057821$ & $-7.515980$ \\
5679  & $-36.904331$ & $30.038645$ & $-0.592389$ & $-0.057820$ & $-7.515896$ \\
7456  & $-36.910731$ & $30.044909$ & $-0.592249$ & $-0.057820$ & $-7.515891$ \\
10304 & $-36.913837$ & $30.047995$ & $-0.592174$ & $-0.057820$ & $-7.515836$ \\ \hline
\multicolumn{6}{|l|}{$^4P_g$} \\ \hline
277  & $-36.062593$ & $29.388974$ & $-0.578999$ & $-0.027932$ & $-7.280549$ \\
406  & $-36.062306$ & $29.387961$ & $-0.576977$ & $-0.027940$ & $-7.279263$ \\
595  & $-36.120148$ & $29.442882$ & $-0.576059$ & $-0.027944$ & $-7.281270$ \\
872  & $-36.135572$ & $29.460497$ & $-0.575635$ & $-0.027944$ & $-7.278654$ \\
1278 & $-36.155086$ & $29.479579$ & $-0.575179$ & $-0.027944$ & $-7.278630$ \\
1873 & $-36.166969$ & $29.491458$ & $-0.574979$ & $-0.027944$ & $-7.278433$ \\
2733 & $-36.172481$ & $29.496815$ & $-0.574859$ & $-0.027943$ & $-7.278469$ \\
3580 & $-36.179044$ & $29.503374$ & $-0.574758$ & $-0.027943$ & $-7.278371$ \\
4672 & $-36.181062$ & $29.505364$ & $-0.574684$ & $-0.027943$ & $-7.278325$ \\ \hline
\multicolumn{6}{|l|}{$^2D_g$} \\ \hline
277  & $-35.970107$ & $29.276227$ & $-0.585496$ & $-0.042998$ & $-7.322373$ \\
406  & $-36.038120$ & $29.340083$ & $-0.583894$ & $-0.043104$ & $-7.325035$ \\
595  & $-36.141397$ & $29.439580$ & $-0.582434$ & $-0.043201$ & $-7.327453$ \\
872  & $-36.182799$ & $29.478829$ & $-0.581222$ & $-0.043269$ & $-7.328461$ \\
1278 & $-36.222970$ & $29.517760$ & $-0.580282$ & $-0.043298$ & $-7.328790$ \\
1873 & $-36.263293$ & $29.557494$ & $-0.579795$ & $-0.043309$ & $-7.328902$ \\
2745 & $-36.281367$ & $29.575076$ & $-0.579327$ & $-0.043314$ & $-7.328933$ \\
4023 & $-36.290237$ & $29.583798$ & $-0.579005$ & $-0.043317$ & $-7.328761$ \\
5858 & $-36.308459$ & $29.601703$ & $-0.578769$ & $-0.043319$ & $-7.328843$ \\
8231 & $-36.313784$ & $29.606950$ & $-0.578644$ & $-0.043319$ & $-7.328797$ \\ \hline
\end{tabular}
\end{table}

In order to compare the computed excitation energies with experimental data,
nuclear mass
proper for particular isotope has to be taken into account. The most abundant 
isotopes of boron are $^{11}B$ and $^{10}B$, whose nuclear masses amount to
20063.7375 a.u. and 18247.4689 a.u., respectively.
The same basis sets are used in the calculations, as for fixed nucleus ---
only the linear parameters are allowed to vary.
Table \ref{tab-finmass} lists the nonrelativistic energies and all components
of scalar relativistic corrections, for the largest basis, for
each state. Extrapolations to complete basis sets are carried out with the
same corrections as for fixed nucleus. Concerning the
terms not appearing for fixed nucleus, $\langle\hat{H}_{1n}\rangle$ is damped
effectively by third power of the nuclear mass in the denominator, and amounts
to few femtohartree only, which is negligible at the accuracy level
achieved in present calculations. On the other hand,
$\langle\hat{H}_{4n}\rangle$ amount to few $\mu$hartree. Other
components' values however change in such extent that total scalar
relativistic corrections differ from those obtained for fixed nucleus
by few nanohartree only.

\begin{table}
\caption{Variationally bound, and extrapolated nonrelativistic energies  (in
hartree), and scalar relativistic corrections (in mhartree) for
$^{11}B$ and $^{10}B$ isotopes of boron \label{tab-finmass}}
\begin{tabular}{|l|c|c|c|c|c|c|} \hline
& $^2P_u$($^{11}B$) & $^2P_u$($^{10}B$) & $^4P_g$($^{11}B$) 
& $^4P_g$($^{10}B$) & $^2D_g$($^{11}B$) & $^2D_g$($^{10}B$) \\ \hline
$E_{nr}$ & $-24.652625854$ & $-24.652502219$ & $-24.520826909$ & $-24.520706030$ & $-24.434765075$ & $-24.434644055$ \\
$E_{extr}$ & $-24.65262669$ & $-24.65250305$ & $-24.52082728$ & $-24.52070640$ & $-24.43476754$ & $-24.43464652$  \\
$\langle\hat{H}_1\rangle$ & $-36.906350$ & $-36.905605$ & $-36.173742$ & $-36.173014$ & $-36.306284$ & $-36.300219$ \\
$\langle\hat{H}_{1n}\rangle$ & $-6.5\cdot 10^{-12}$ & $-8.6\cdot 10^{-12}$ & $-6.3\cdot 10^{-12}$ & $-8.4\cdot 10^{-12}$ & $-6.3\cdot 10^{-12}$ & $-8.4\cdot 10^{-12}$ \\
$\langle\hat{H}_2\rangle$ & $30.043429$ & $30.042975$ & $29.500886$ & $29.500440$ & $29.602345$ & $29.596644$ \\
$\langle\hat{H}_3\rangle$ & $-0.592094$ & $-0.592086$ & $-0.574607$ & $-0.574600$ & $-0.578563$ & $-0.578680$ \\
$\langle\hat{H}_4\rangle$ & $-0.057750$ & $-0.057743$ & $-0.027879$ & $-0.027873$ & $-0.043248$ & $-0.043241$ \\
$\langle\hat{H}_{4n}\rangle$ & $-0.003040$ & $-0.003342$ & $-0.002961$ & $-0.003255$ & $-0.002977$ & $-0.003273$ \\
$E_{RS}$ & $-7.515805$ & $-7.515802$ & $-7.278303$ & $-7.278301$ & $-7.328727$ & $-7.328720$ \\\hline
\end{tabular}
\end{table}

The wavenumbers proper for excitations from the ground state to the lowest
$^4P_g$ and $^2D_g$ states, calculated for $^{11}B$, and not accountig for
the fine structure, amount to
$28978.75$~cm$^{-1}$ and $47855.62$~cm$^{-1}$, respectively.
The latter differs from the experiment-based one by 9~cm$^{-1}$, which is
comparable with the energy difference between the $^2P_{1/2}$ and
$^2P_{3/2}$ states (fine structure, 15 cm$^{-1}$) \cite{ASD}.
Similar accuracy is expected for the excitation energy to the $^4P_g$ state.

The isotopic shifts may be easily calculated from present results.
The differences of term energies, between $^{11}B$ and $^{10}B$, computed with
the same basis,
remain very stable as the basis size is increased -- similarly as
for the carbon atom \cite{C-KS}. They are given in table \ref{mshift},
with larger number of significant digits than total energy, for
two largest basis sets.
Isotopic shift of $-0.57316$ cm$^{-1}$ is obtained for the
$^2P_u\rightarrow^2D_g$ excitiation, while the measured
value, averaged over two spectral lines, is equal to $-0.569(3)$ cm$^{-1}$ 
\cite{ishift}. $-0.60502$~cm$^{-1}$ is predicted for the
$^2P_u\rightarrow^4P_g$ transitions.

\begin{table}
\caption{Isotopic shifts for term energies (components in hartree, total in
cm$^{-1}$) \label{mshift}}
\begin{tabular}{|r|c|c|c|} \hline
 K    & $E_{nr}(^{10}B)-E_{nr}(^{11}B$  & $E_{RS}(^{10}B)-E_{RS}(^{11}B)$ & $E_{rel}(^{10}B)-E_{rel}(^{11}B)$  \\ \hline
\multicolumn{4}{|l|}{$^2P_u$} \\ \hline
7456  & $0.0001236349$ & $3.0\cdot 10^{-9}$ & 27.13538 \\
10304 & $0.0001236348$ & $3.1\cdot 10^{-9}$ & 27.13538 \\ \hline
\multicolumn{4}{|l|}{$^4P_g$} \\ \hline
3580  & $0.0001208790$ & $2.2\cdot 10^{-9}$ & 26.53036 \\
4672  & $0.0001208790$ & $2.2\cdot 10^{-9}$ & 26.53036 \\ \hline
\multicolumn{4}{|l|}{$^2D_g$} \\ \hline
5858  & $0.0001210195$ & $6.8\cdot 10^{-9}$ & 26.56220 \\
8231  & $0.0001210195$ & $6.9\cdot 10^{-9}$ & 26.56222 \\ \hline
\end{tabular}
\end{table}

\section{Conclusions}

The present work provides most accurate to date, nonrelativistic
energies of the lowest $^2P_u$, $^4P_g$ abd $^2D_g$ states of the boron atom.
With scalar relativistic
corrections and finite nuclear mass taken into account, term energies are
obtained, whose main source of remaining error is the missing fine structure.
The measured fine splitting amounts to c.a. 15 cm$^{-1}$ for the $^2P_u$ term,
c.a. 11 cm$^{-1}$ for the $^4P_g$ term, and less than 1~cm$^{-1}$ for the
$^2D_g$ term \cite{ASD}. The computed $^2P_u\rightarrow^2D_g$ excitation
energy confirms the experiment-based result within c.a. 11 cm$^{-1}$, and
comparable accuracy is expected for the $^2P_u\rightarrow^4P_g$ excitation.
This reveals gross inaccuracy of the
latter excitation energy, based on experimental data for
heavier, isoelectronic ions. This inaccuracy exceeds 300 cm$^{-1}$,
therefore an update of the content of
Atomic Spectra Database \cite{ASD} would be recommended, concerning the
energies of the quartet states of boron atom.
It is worth noting that the predictions of the MCHF study \cite{CFF} were
accurate within 20~cm$^{-1}$.
Further calculations that would include splitting of energy
levels due to magnetic spin-orbit and spin-spin couplings are desired.

On technical side of the work, it is proven again that the symmetrized,
explicitly correlated Gaussian lobe functions form an efficient basis
for atomic states,
in spite of not being eigenfunctions of the $\hat{L}^2$ operator.
Lower variational energies are obtained at shorter expansions, than with
basis functions having exact symmetry properties.

\noindent
{\bf Acknowledgments}

This research was supported by Department of Physical and Quantum Chemistry
of Wroc{\l}aw University of Science and Technology. Most calculations have
been carried out in Wroc{\l}aw Center for Networking and Supercomputing
(WCSS, http://wcss.pl).

\end{document}